\documentclass[a4paper,12pt]{article}
\usepackage[top=3cm, bottom=3cm, left=3cm,right=3cm]{geometry}
\usepackage{amssymb}
\usepackage{subfigure}
\usepackage{graphicx}
\usepackage{epsfig}
\usepackage{booktabs}
\usepackage{array}
\usepackage{multirow}
\usepackage{cite}
\usepackage{appendix}
\usepackage{amsmath}
\usepackage{hyperref}
\usepackage{setspace}

\begin{document}
\title{A lower bound on the Longitudinal Structure Function at small \textit{x} from a self-similarity based model of Proton}
\author{$ \mathrm{Akbari \; Jahan}^{\star} $ and D K Choudhury \\ Department of Physics, Gauhati University,\\ Guwahati - 781 014, Assam, India.\\ $ {}^{\star} $Email: akbari.jahan@gmail.com}
\date{}
\maketitle
\begin{abstract}
Self-similarity based model of proton structure function at small \textit{x} was reported in the literature sometime back. The phenomenological validity of the model is in the kinematical region $ 6.2\, \times \, 10^{-7} \leq x \leq 10^{-2}$ and $ 0.045 \leq Q^{2} \leq 120 \, \mathrm{GeV^{2}} $. We use momentum sum rule to pin down the corresponding self-similarity based gluon distribution function valid in the same kinematical region. The model is then used to compute bound on the longitudinal structure function $F_{L}\left(x,Q^{2} \right)$ for Altarelli-Martinelli equation in QCD and is compared with the recent HERA data.
\textbf{Keywords \,:} Self-similarity, small \textit{x}, longitudinal structure function, QCD.
\end{abstract}

\section{Introduction}
Sometime back \cite{1}, we reported results of the longitudinal structure function $F_{L}\left(x,Q^{2} \right)$ at small \textit{x} using a self-similarity based model of quark and gluon distributions \cite{2,3}. The model was alternative to the model of proton suggested by Lastovicka \cite{4}. We did not pursue the model of Ref.\cite{4} in \cite{1} due to the presence of a singularity at $ x \approx 0.019 $ (which lies outside its range of validity) although the longitudinal structure function involves an integration upto $x = 1 $. The limitation of the model of Ref.\cite{2,3} is that it did not specify the range of its validity unlike Ref.\cite{4} and hence qualitatively less reliable. In Ref.\cite{1}, we therefore assumed its validity in the entire \textit{x} range to estimate the longitudinal structure function.\\

In the present work, we use the model of Ref.\cite{4} instead. We invoke momentum sum rule \cite{5} in the range $ x_{a}\leq x \leq x_{b} $ (where $x_{a}= 6.2\, \times \, 10^{-7}$ and $ x_{b}=10^{-2} $), compute the momentum fraction of quarks and gluons and pin down the gluon distribution numerically. We then use it to compute the bound on the longitudinal structure function $F_{L}\left(x,Q^{2} \right)$ for Altarelli-Martinelli equation in QCD \cite{6} and compare it using recent data \cite{7}.\\

In Sec. 2 we outline the formalism, and Sec. 3 gives the results and discussion. The conclusions of this work are highlighted in Sec. 4.
\section{Formalism}
\subsection{Self-similarity based model of proton structure function and the momentum sum rule inequality}
In Ref.\cite{4}, the following form of parton density function (PDF) and the structure function $F_{2}\left(x,Q^{2} \right)$ were suggested based on the notion of self-similarity,
\begin{eqnarray}
\label{eqn:qi}
q_{i}\left(x,Q^{2}\right) & = & \frac{1}{M^{2}} \frac{e^{D_{0}^{i}}\, Q_{0}^{2} \, x^{-D_{2}}}{1+D_{3}+D_{1} \log \left(\frac{1}{x} \right)} \, \left( \left( \frac{1}{x}\right)^{D_{1} \log \left(1+ \frac{Q^{2}}{Q_{0}^{2}} \right)} \, \left(1+ \frac{Q^{2}}{Q_{0}^{2}} \right)^{D_{3}+1} -1 \right) \nonumber \\
\end{eqnarray}
\begin{eqnarray}
\label{eqn:F2}
F_{2}\left(x,Q^{2}\right) & = & \frac{1}{M^{2}} \frac{e^{D_{0}}\, Q_{0}^{2} \, x^{-D_{2}+1}}{1+D_{3}+D_{1} \log \left(\frac{1}{x} \right)} \, \left( \left( \frac{1}{x}\right)^{D_{1} \log \left(1+ \frac{Q^{2}}{Q_{0}^{2}} \right)} \, \left(1+ \frac{Q^{2}}{Q_{0}^{2}} \right)^{D_{3}+1} -1 \right) \nonumber \\
\end{eqnarray}
valid in the range $ 6.2\, \times \, 10^{-7} \leq x \leq 10^{-2}$ and $ 0.045 \leq Q^{2} \leq 120 \, \mathrm{GeV^{2}} $. The mass scale $ M^{2}$ was set to be 1 $\mathrm{GeV^{2}}$. Here, $ e^{D_{0}}=\sum \limits_{i=1}^{n_{f}} e_{i}^{2}\left( e^{D_{0}^{i}}+ e^{\bar{D_{0}^{i}}}\right) $. The parameters are \cite{4}
\begin{eqnarray}
\label{eqn:parameters}
D_{0} & = & 0.339 \pm 0.145 \nonumber \\
D_{1} & = & 0.073\pm 0.001 \nonumber \\
D_{2} & = & 1.013\pm 0.01 \nonumber \\
D_{3} & = & -1.287\pm 0.01 \nonumber \\
Q_{0}^{2} & = & 0.062\pm 0.01 \; \mathrm{GeV}^{2}
\end{eqnarray}
The formalism of momentum sum rule \cite{5} has been reported recently in Ref.\cite{8}. We briefly outline here for completeness.\\

The momentum sum rule inequality yields \cite{8}
\begin{equation}
\int\limits_{x_{a}}^{x_{b}}\lbrace a \, F_{2}\left( x,Q^{2}\right) + G\left( x,Q^{2}\right) \rbrace \, dx \leq 1
\end{equation}
where $\displaystyle a=\frac{e^{\tilde{D_{0}}}}{e^{D_{0}}}$ with $ e^{\tilde D_{0}}=\sum \limits_{i=1}^{n_{f}} \left( e^{D_{0}^{i}}+ e^{\bar{D_{0}^{i}}}\right) $. The parameter \textit{a} is \textit{x} independent. Its value has been determined from data in our earlier paper \cite{9} and is found to be approximately equal to 3.1418 (for fractional charged quarks). For integral charged partons, \textit{a} = 1 \cite{10}. We define the total $ \left( 0 \leq x \leq 1 \right)$ and the partial $ \left( x_{a} \leq x \leq x_{b} \right)$ momentum fractions of quarks and gluons respectively as $\langle x \rangle_{q}$, $\langle \hat x \rangle_{q}$, $\langle x \rangle_{g}$ and $ \langle \hat x \rangle_{g}$ \cite{8}. We have seen that one can evaluate $\langle \hat x \rangle_{q}$ for any $Q^{2}$, which is given by the relation \cite{8}:
\begin{equation}
\label{eqn:Q2}
Q^{2} = Q_{0}^{2}\left\lbrace exp \left(\frac{\langle \hat x \rangle_{q}}{a \, e^{D_{0}}} \, \frac{M^{2}}{Q_{0}^{2}} \, \frac{e^{-\left(\frac{D_{3}+1}{D_{1}} \right) \left(2-D_{2} \right)}}{\log \left(\frac{x_{b}}{x_{a}}\right)} \right)-1   \right\rbrace
\end{equation}
Using Eq. (\ref{eqn:Q2}), we calculate the numerical values of $\langle \hat x \rangle_{q}$ for a few representative values of $Q^{2}$ and subsequently obtain the maximum limit of $\langle \hat x \rangle_{g}$. We then tabulate them in Table \ref{tab:hQ2}. We observe that as $Q^{2}$ increases, $\langle \hat x \rangle_{q}$ also increases but $\langle \hat x \rangle_{g}$ decreases. We also note that the partial momentum fractions of quarks ($\langle \hat x \rangle_{q}$) and gluons ($\langle \hat x \rangle_{g}$) vary with $Q^{2}$ and thus can be related as
\begin{equation}
\label{eqn:xghat_xqhathQ2}
\langle \hat x \rangle_{g} \approx h \left(Q^{2} \right)\, \langle \hat x \rangle_{q}
\end{equation}
where $h \left(Q^{2} \right)$ is any $Q^{2}$ dependent function and it decreases with $Q^{2}$ in the GeV scale (Table \ref{tab:hQ2}).

\begin{table}[!h]
\begin{center}
\caption[Values of $\langle \hat x \rangle_{q}$, $\langle \hat x \rangle_{g}$ and $h \left(Q^{2} \right)$ for given values of $Q^{2}$.]{\textbf{Values of $\langle \hat x \rangle_{q}$, $\langle \hat x \rangle_{g}$ and $h \left(Q^{2} \right)$ for given values of $Q^{2}$.}}
\label{tab:hQ2}
\medskip
\begin{tabular}{|c|c|c||c|c||c|c|}
\hline
$Q^{2} \left(\mathrm{GeV^{2}} \right) $ & \multicolumn{2}{c}{$\langle \hat x \rangle_{q}$} \vline & \multicolumn{2}{c}{$\langle \hat x \rangle_{g}$} \vline &  \multicolumn{2}{c}{$h \left(Q^{2} \right)$} \vline \\ \cline{2-3} \cline{4-5} \cline{6-7}
{} & $a=1$   & $a=3.1418$ & $a=1$   & $a=3.1418$ & $a=1$   & $a=3.1418$ \\
\hline
10 & 0.0886 & 0.2783 & 0.9114 & 0.7217 & 10.2902 & 2.5935 \\
20 & 0.1006 & 0.3160 & 0.8994 & 0.6840 & 8.9421 & 2.1645 \\
30 & 0.1076 & 0.3381 & 0.8924 & 0.6619 & 8.2919 & 1.9575 \\
40 & 0.1126 & 0.3538 & 0.8874 & 0.6462 & 7.8796 & 1.8263 \\
50 & 0.1165 & 0.3660 & 0.8835 & 0.6340 & 7.5840 & 1.7322 \\
60 & 0.1197 & 0.3760 & 0.8803 & 0.6240 & 7.3568 & 1.6598 \\
70 & 0.1223 & 0.3844 & 0.8776 & 0.6156 & 7.1736 & 1.6016 \\
80 & 0.1247 & 0.3917 & 0.8753 & 0.6083 & 7.0214 & 1.5531 \\
90 & 0.1267 & 0.3981 & 0.8733 & 0.6019 & 6.8917 & 1.5118 \\
\hline
\end{tabular}
\end{center}
\end{table}

\subsection{Self-similarity based gluon distribution function}
It is important to know the gluon distribution inside a hadron at small \textit{x} because gluons are expected to be dominant in this region. The steep increase in $F_{2} \left(x,Q^{2} \right)$ towards small \textit{x} observed at HERA also indicates a similar increase in gluon distribution towards small \textit{x} in perturbative QCD. Accurate knowledge of gluon distribution function at small \textit{x} and large virtuality $Q^{2}$ plays a vital role in estimating QCD backgrounds and in calculating gluon-initiated processes, and thus in our ability to search for new physics at the LHC. The gluon and quark distribution functions have traditionally been determined simultaneously by fitting experimental data on neutral and charged current deep inelastic scattering processes and some jet data over a large domain of values of \textit{x} and $Q^{2}$. The distributions at small \textit{x} and large $Q^{2}$ are determined mainly by the proton structure function $F_{2} \left(x,Q^{2} \right)$ measured in deep inelastic \textit{ep} scattering \cite{11}.\\

The exact relation between the gluon distribution $G\left(x,Q^{2} \right)=xg\left(x,Q^{2} \right)$ and the quark distribution $F_{2}\left(x,Q^{2} \right)= x \sum \limits_{i} e_{i}^{2}  \left\lbrace q_{i}\left(x,Q^{2} \right)+ \bar{q}_{i}\left(x,Q^{2} \right) \right\rbrace$ is not derivable in QCD even in leading order (LO). However, simple forms of such relation are available in literature to facilitate the analytical solutions of coupled DGLAP equations. In Ref.\cite{12}, it was assumed that $Q^{2}$ dependence of both the distributions are identical. In Ref.\cite{13}, on the other hand, the following simple relation was assumed:
\begin{equation}
\label{eqn:JKS}
G \left(x,Q^{2} \right) = k \, F_{2}\left(x,Q^{2} \right)
\end{equation}
where \textit{k} is a parameter to be determined from experiments.\\

A more rigorous analysis was done by Lopez and Yndurain \cite{14} and they investigated the behavior of the (singlet) structure function $F_{2}\left(x,Q^{2} \right)$ as $x \rightarrow 0$, under the assumption that it is of power type (with eventually logarithmic) and working in leading order (LO). It was observed that for $x \rightarrow 0$ (Eq. (3.15a) and Eq. (3.15b) of Ref.\cite{14}):
\begin{equation}
\label{eqn:F2_Lopez}
F_{2}\left(x,Q^{2} \right) \underset{x \rightarrow 0}{\simeq} B_{S} \left[\alpha_{s}(Q^{2})\right]^{-d_{+}\left(1+\lambda_{S} \right)} x^{- \lambda_{S}}
\end{equation}
and
\begin{equation}
\label{eqn:G_Lopez}
G\left(x,Q^{2} \right) \underset{x \rightarrow 0}{\simeq} B_{G} \left[\alpha_{s}(Q^{2})\right]^{-d_{+}\left(1+\lambda_{S} \right)} x^{- \lambda_{S}}
\end{equation}
where the functions $B_{S}$ and $B_{G}$ are $Q^{2}$ dependent. $d_{+}\left(1+\lambda_{S} \right)$ is the largest eigenvalue of the anomalous dimension matrix (Eq. (1.3b) of Ref.\cite{14}) and $\lambda_{S}$ is strictly positive. From Eqs. (\ref{eqn:F2_Lopez}) and (\ref{eqn:G_Lopez}), one infers that the ratio of the gluon and the (singlet) structure function is only $Q^{2}$ dependent. That is,
\begin{equation}
\label{eqn:GF2_ratio}
\frac{G\left(x,Q^{2} \right)}{F_{2}\left(x,Q^{2} \right)} \, \underset{x \rightarrow 0}{\simeq} \, \frac{B_{G}}{B_{S}} \, \simeq \, f(Q^{2})
\end{equation}
In Table \ref{tab:hQ2}, we have already observed that there is a relative $Q^{2}$ dependence between $\langle \hat x \rangle_{q}$ and $\langle \hat x \rangle_{g}$. A simple plausible way of realizing it is by assuming a relation
\begin{equation}
\label{eqn:xg}
x g\left(x,Q^{2} \right)= c \left(Q^{2}\right) x \sum_{i}\left\lbrace q_{i}\left( x,Q^{2}\right)+ \bar{q_{i}}\left( x,Q^{2}\right) \right\rbrace
\end{equation}
or equivalently,
\begin{equation}
\label{eqn:G}
G\left(x,Q^{2} \right)= c \left(Q^{2}\right) a \, F_{2}\left(x,Q^{2} \right)
\end{equation}
where $c \left(Q^{2}\right)$ is a $Q^{2}$ dependent function which is also compatible with Eq. (\ref{eqn:GF2_ratio}). Eqs. (\ref{eqn:xg}) and (\ref{eqn:G}) therefore imply that
\begin{equation}
\langle x \rangle_{g} = c \left(Q^{2}\right) \langle x \rangle_{q}
\end{equation}
and
\begin{equation}
\langle \hat x \rangle_{g} = c \left(Q^{2}\right) \langle \hat x \rangle_{q}
\end{equation}
leading to
\begin{equation}
\label{eqn:hc}
h \left(Q^{2} \right) \equiv c \left(Q^{2} \right)
\end{equation}
from Eq. (\ref{eqn:xghat_xqhathQ2}).

\subsection{Longitudinal structure function at small \textit{x}}
In QCD, the longitudinal structure function $F_{L}\left(x,Q^{2} \right)$ is given by the Altarelli-Martinelli equation \cite{6}:
\begin{small}
\begin{equation}
\label{eqn:AM}
F_{L}\left(x,Q^{2} \right)=\frac{\alpha_{s}\left(Q^{2}\right)}{\pi}\left[\frac{4}{3} \int \limits_{x}^{1}\frac{dy}{y} \left(\frac{x}{y}\right)^{2} F_{2}\left(y,Q^{2} \right)+ 2 \sum_{i}e_{i}^{2} \int \limits_{x}^{1}\frac{dy}{y}\left(\frac{x}{y}\right)^{2} \left(1-\frac{x}{y} \right) G \left(y,Q^{2} \right) \right]
\end{equation}
\end{small}
where $e_{i}$ is the electric charge of the \textit{i}th parton and $\alpha_{s}$ is the strong coupling constant. Considering only the leading order and the number of flavors, $n_{f}= 4 $, $\alpha_{s}$ is given by the following relation \cite{15}:\\
\begin{equation}
\alpha_{s} \left(Q^{2}\right)=\frac{12 \pi}{25 \ln \left(\frac{Q^{2}}{\Lambda^{2}} \right)}
\end{equation}
where $ \Lambda$ is the QCD scale. We take $\Lambda_{\bar{MS}}^{(4)} = 296 \pm 10 $ MeV \cite{15}.\\

In the limited range $ x_{a} \leq x \leq x_{b}$, Altarelli-Martinelli equation yields lower bound on $F_{L}\left(x,Q^{2} \right)$, say $\hat{F}_{L}\left(x,Q^{2} \right)$. That is,\\
\begin{small}
\begin{equation}
\label{eqn:AM_bound}
\hat{F}_{L}\left(x,Q^{2} \right)=\frac{\alpha_{s}\left(Q^{2}\right)}{\pi}\left[\frac{4}{3} \int \limits_{x}^{x_{b}}\frac{\mathrm{d} y}{y} \left(\frac{x}{y}\right)^{2} F_{2}\left(y,Q^{2} \right)+ 2 \sum_{i}e_{i}^{2} \int \limits_{x}^{x_{b}}\frac{\mathrm{d}y}{y}\left(\frac{x}{y}\right)^{2} \left(1-\frac{x}{y} \right) G \left(y,Q^{2} \right) \right]
\end{equation}
\end{small}
Using Eqs. (\ref{eqn:G}) and (\ref{eqn:hc}) in Eq. (\ref{eqn:AM_bound}), we get the expression for $ \hat{F}_{L}\left(x,Q^{2} \right) $ in terms of the proton structure function only. It implies\\

$\hat{F}_{L}\left(x,Q^{2} \right) =$
\begin{equation}
\label{eqn:AM_FLbound}
\frac{\alpha_{s}\left(Q^{2}\right)}{\pi} \left[\frac{4}{3} \int \limits_{x}^{x_{b}}\frac{\mathrm{d}y}{y} \left(\frac{x}{y}\right)^{2} F_{2}\left(y,Q^{2} \right)+ 2\, a \, h \left(Q^{2}\right) \sum_{i}e_{i}^{2} \int \limits_{x}^{x_{b}}\frac{\mathrm{d}y}{y}\left(\frac{x}{y}\right)^{2} \left(1-\frac{x}{y} \right) F_{2} \left(y,Q^{2} \right) \right]
\end{equation}
Consider
\begin{equation}
\label{eqn:I1_int}
I_{1}\left(x,Q^{2} \right) = \int \limits_{x}^{x_{b}}\frac{dy}{y} \left(\frac{x}{y}\right)^{2} F_{2}\left(y,Q^{2} \right)
\end{equation}
\begin{equation}
\label{eqn:I2_int}
I_{2}\left(x,Q^{2} \right) = \int \limits_{x}^{x_{b}}\frac{dy}{y} \left(\frac{x}{y}\right)^{2} \left(1-\frac{x}{y} \right) F_{2}\left(y,Q^{2} \right)
\end{equation}
Using the expression of structure function from Eq. (2) in the above two integration (Eqs. (\ref{eqn:I1_int}) and (\ref{eqn:I2_int})) and applying suitable change of variables as was done in our recent work \cite{8}, we obtain two integrals in each case which are infinite series of the form \cite{16,17}
\begin{equation}
\label{eqn:series}
\int \frac{e^{\mu \, z}}{z} \, dz = \log \vert z \vert + \sum_{n=1}^{\infty} \frac{\mu^{n} \, z^{n}}{n.n!}
\end{equation}
Taking only the first two terms of the series (\ref{eqn:series}), $ I_{1}$ and $ I_{2}$ become
\begin{equation}
I_{1}\left(x,Q^{2} \right) = -x^{2} \, \frac{e^{D_{0}}}{D_{1}} \, \frac{Q_{0}^{2}}{M^{2}} \, e^{-\left(D_{2}+1 \right)\left(\frac{D_{3}+1}{D_{1}} \right)} \, D_{1} \log
\left(1+ \frac{Q^{2}}{Q_{0}^{2}} \right) \, \log \left(\frac{x}{x_{b}} \right)
\end{equation}
and
\begin{equation}
I_{2}\left(x,Q^{2} \right) = I_{1}\left(x,Q^{2} \right)+ x^{3} \, \frac{e^{D_{0}}}{D_{1}} \, \frac{Q_{0}^{2}}{M^{2}} \, e^{-\left(D_{2}+2 \right)\left(\frac{D_{3}+1}{D_{1}} \right)} \, D_{1} \log
\left(1+ \frac{Q^{2}}{Q_{0}^{2}} \right) \, \log \left(\frac{x}{x_{b}} \right)
\end{equation}

Thus calculating the integral on the RHS of Eq. (\ref{eqn:AM_FLbound}), we obtain the final expression of $\hat{F}_{L}\left(x,Q^{2} \right)$ as:\\
\begin{equation}
\label{eqn:FLbound}
\hat{F}_{L}\left(x,Q^{2} \right)=\frac{\alpha_{s}\left(Q^{2}\right)}{\pi} \left[\frac{4}{3} \, I_{1}\left(x,Q^{2} \right)+ \frac{20}{9} \, a \, h \left(Q^{2}\right) \, I_{2}\left(x,Q^{2} \right) \right]
\end{equation}

In Table \ref{tab:FL_bound}, we list the values of the longitudinal structure function $ F_{L}\left(x,Q^{2} \right)$ obtained from experimental data \cite{7} as well as the values of $\hat{F}_{L}\left(x,Q^{2} \right)$ as obtained from our analytical calculations.
\begin{table}[!h]
\begin{center}
\caption{\textbf{The measured $ F_{L}\left(x,Q^{2} \right)$ and the calculated $\hat{F}_{L}\left(x,Q^{2} \right)$ for given values of \textit{x} and $Q^{2}$.}}
\label{tab:FL_bound}
\bigskip
\begin{tabular}{|c|c|c|c|c|} 
\hline
$Q^{2} \left(\mathrm{GeV^{2}} \right) $ & \textit{x} & $ F_{L}\left(x,Q^{2} \right)$ (Ref.\cite{7}) & \multicolumn{2}{c}{$\hat{F}_{L}\left(x,Q^{2} \right)$} \vline \\ \cline{4-5}
{} & {} & {} & $a=1$   & $a=3.1418$ \\
\hline
12 & 0.00028 & 0.22$\pm$ 0.11 & 0.00079 & 0.00063 \\
15 & 0.00037 & 0.08$\pm$ 0.11 & 0.00121 & 0.00095 \\
20 & 0.00049 & 0.24$\pm$ 0.10 & 0.00182 & 0.00141 \\
25 & 0.00062 & 0.38$\pm$ 0.10 & 0.00256 & 0.00197 \\
35 & 0.00093 & 0.24$\pm$ 0.13 & 0.00457 & 0.00345 \\
45 & 0.0014 & 0.18$\pm$ 0.18 & 0.00803 & 0.00597 \\
60 & 0.0022 & 0.33$\pm$ 0.27 & 0.01397 & 0.01025 \\
90 & 0.0036 & 0.48$\pm$ 0.39 & 0.02198 & 0.01580 \\
\hline
\end{tabular}
\end{center}
\end{table}

\section{Results and Discussion}
\begin{figure}[h]
%\centering
\begin{minipage}{18pc}
\includegraphics[scale=.42]{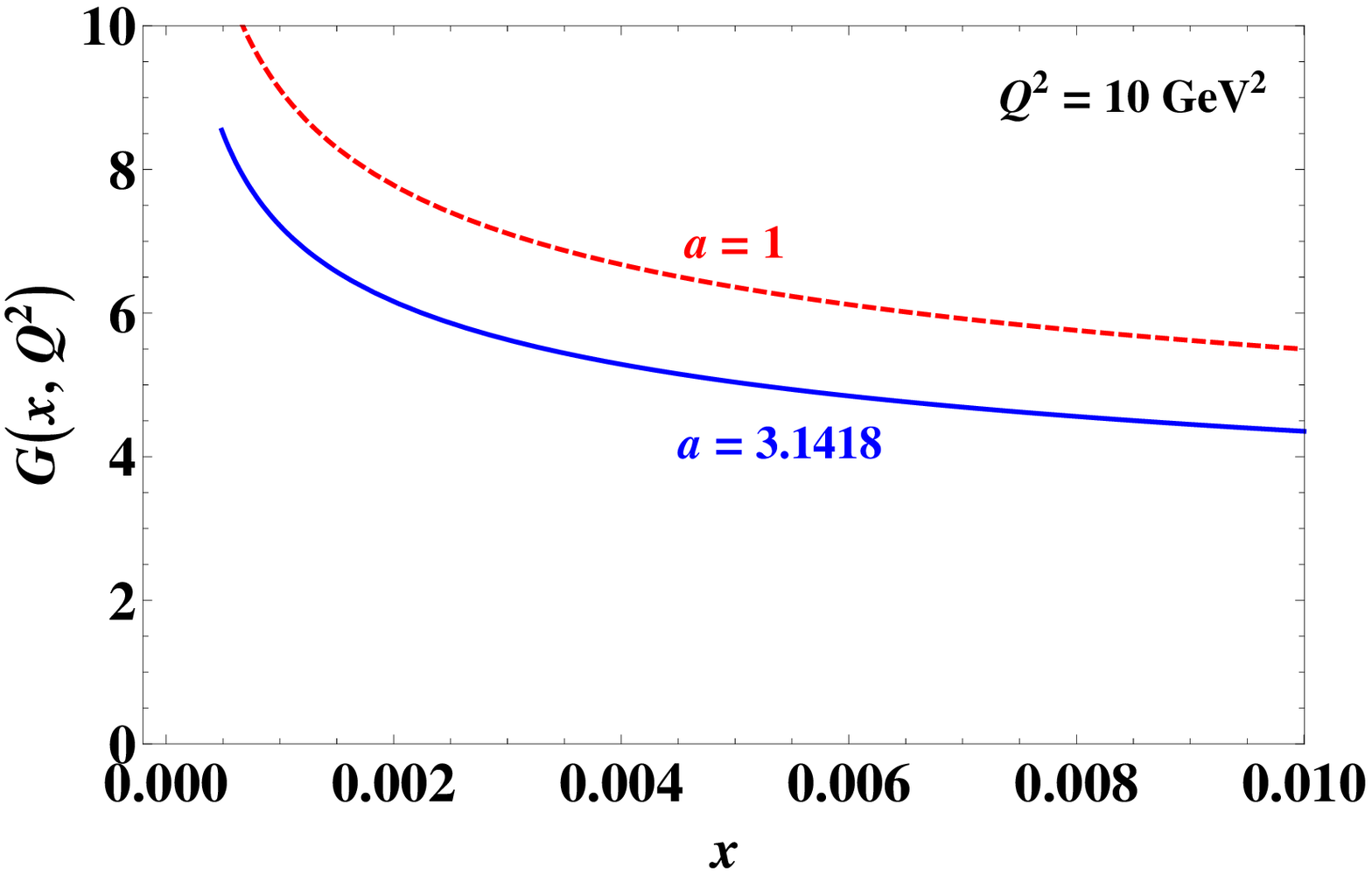}\\
\quad\\
\includegraphics[scale=.42]{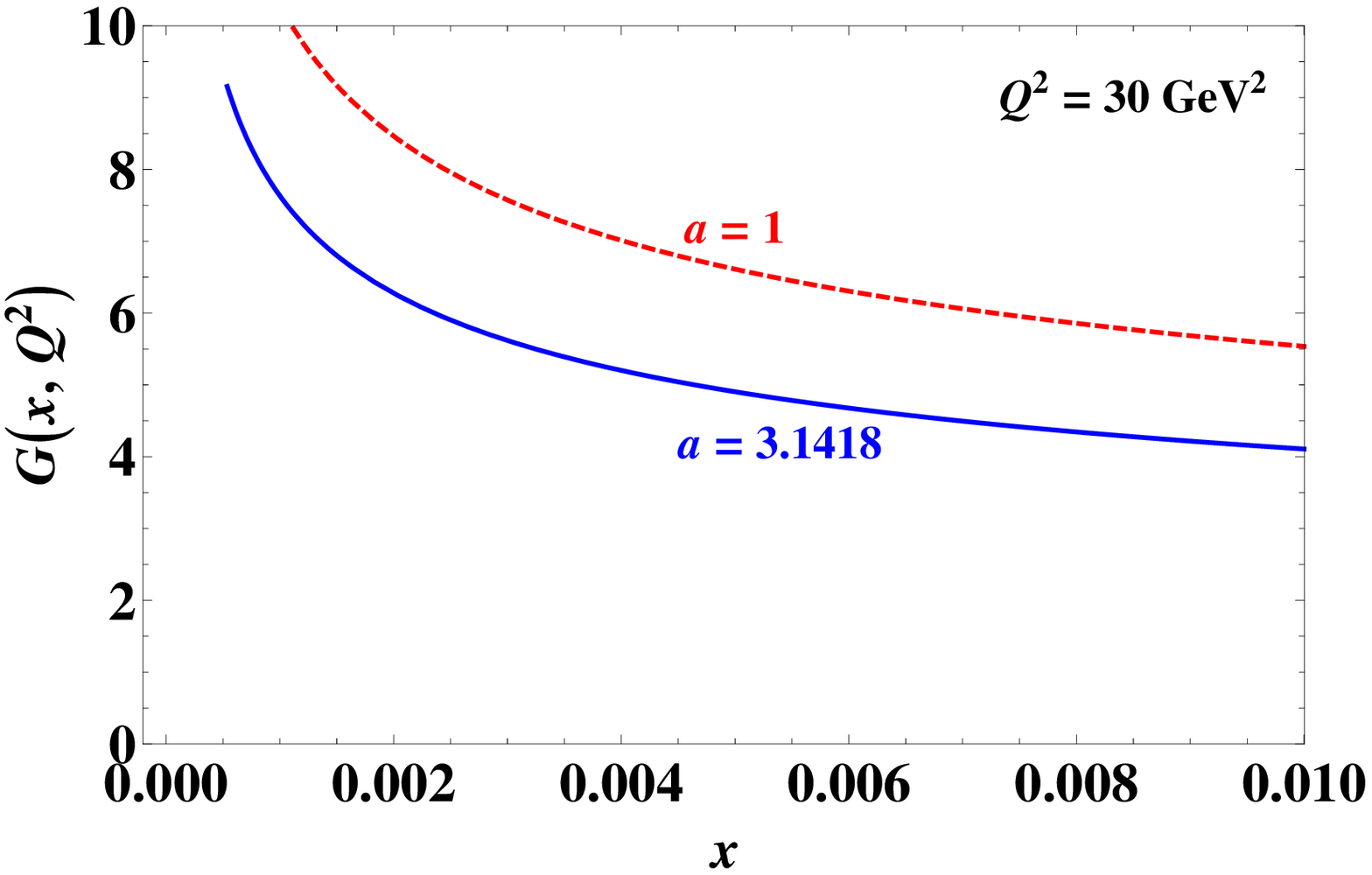}\\
\quad\\
\includegraphics[scale=.42]{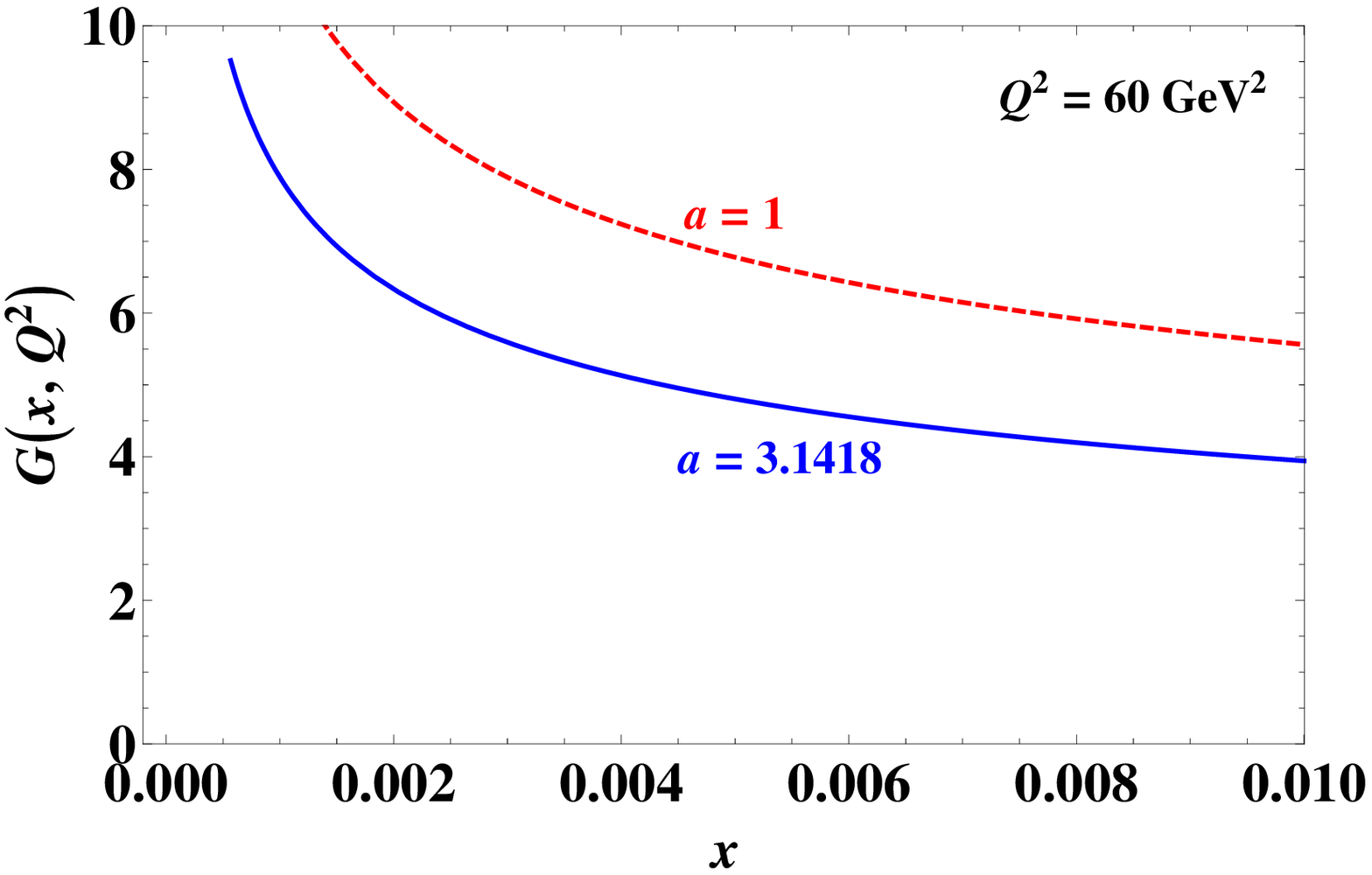}
\caption{The gluon density $G\left(x,Q^{2} \right)$ shown as a function of $ x $ at the given values of $Q^{2}$. The graphs show that the gluon density increases with $Q^{2}$.}
\label{fig:G1}
\end{minipage} \hspace{1pc}%
\begin{minipage}{18pc}
\vspace{0.15in}
\includegraphics[scale=.42]{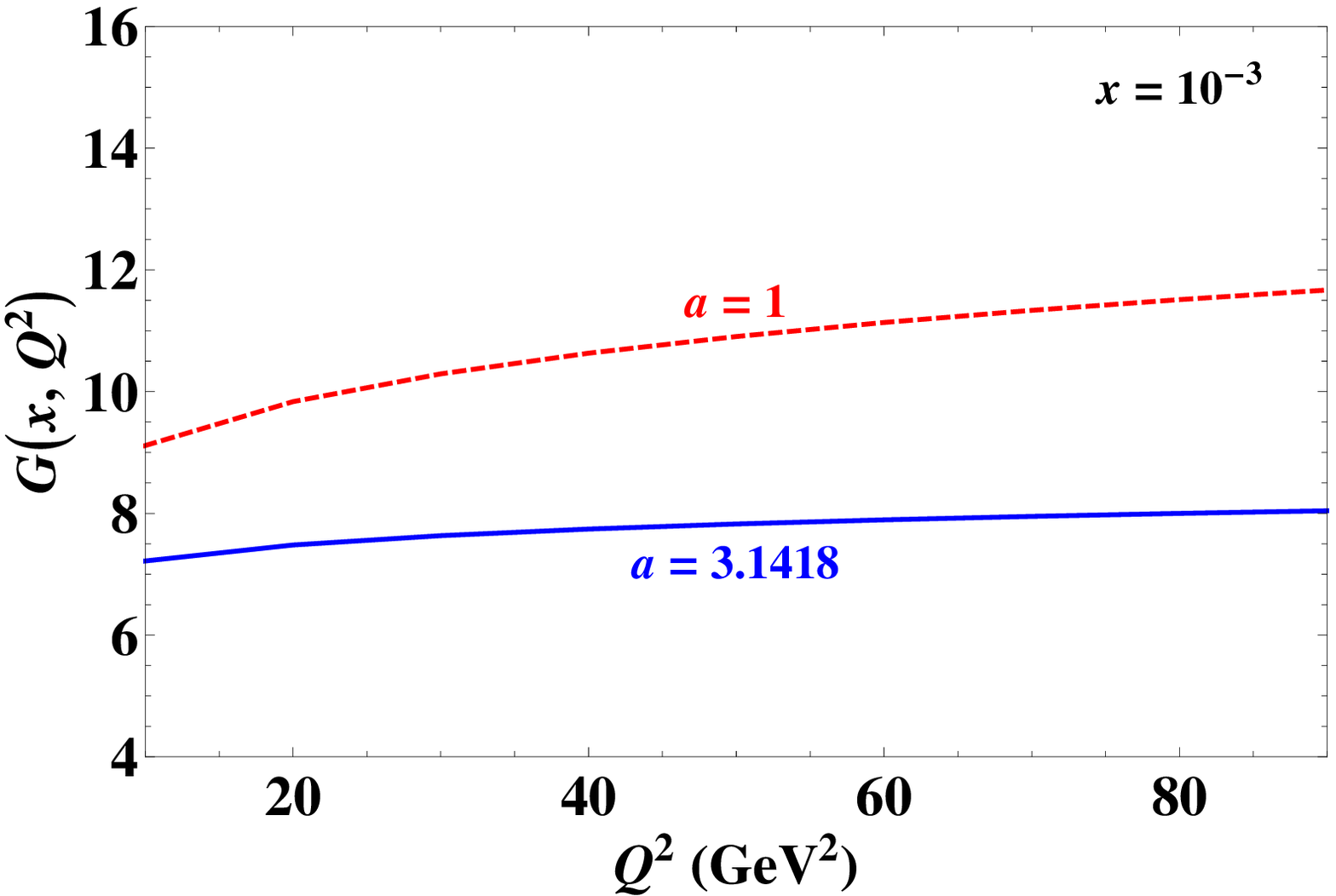}\\
\quad\\
\includegraphics[scale=.42]{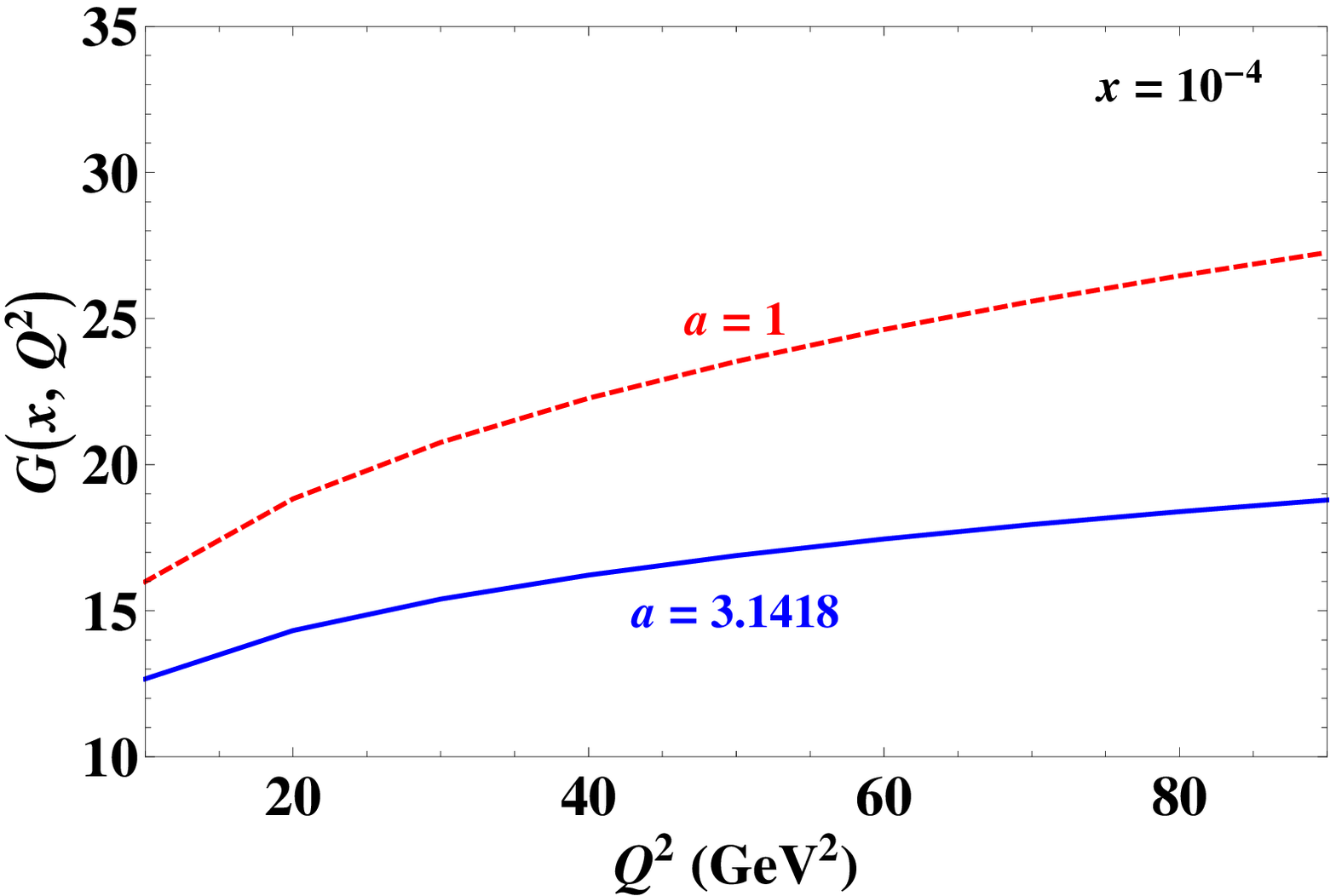}\\
\quad\\
\includegraphics[scale=.42]{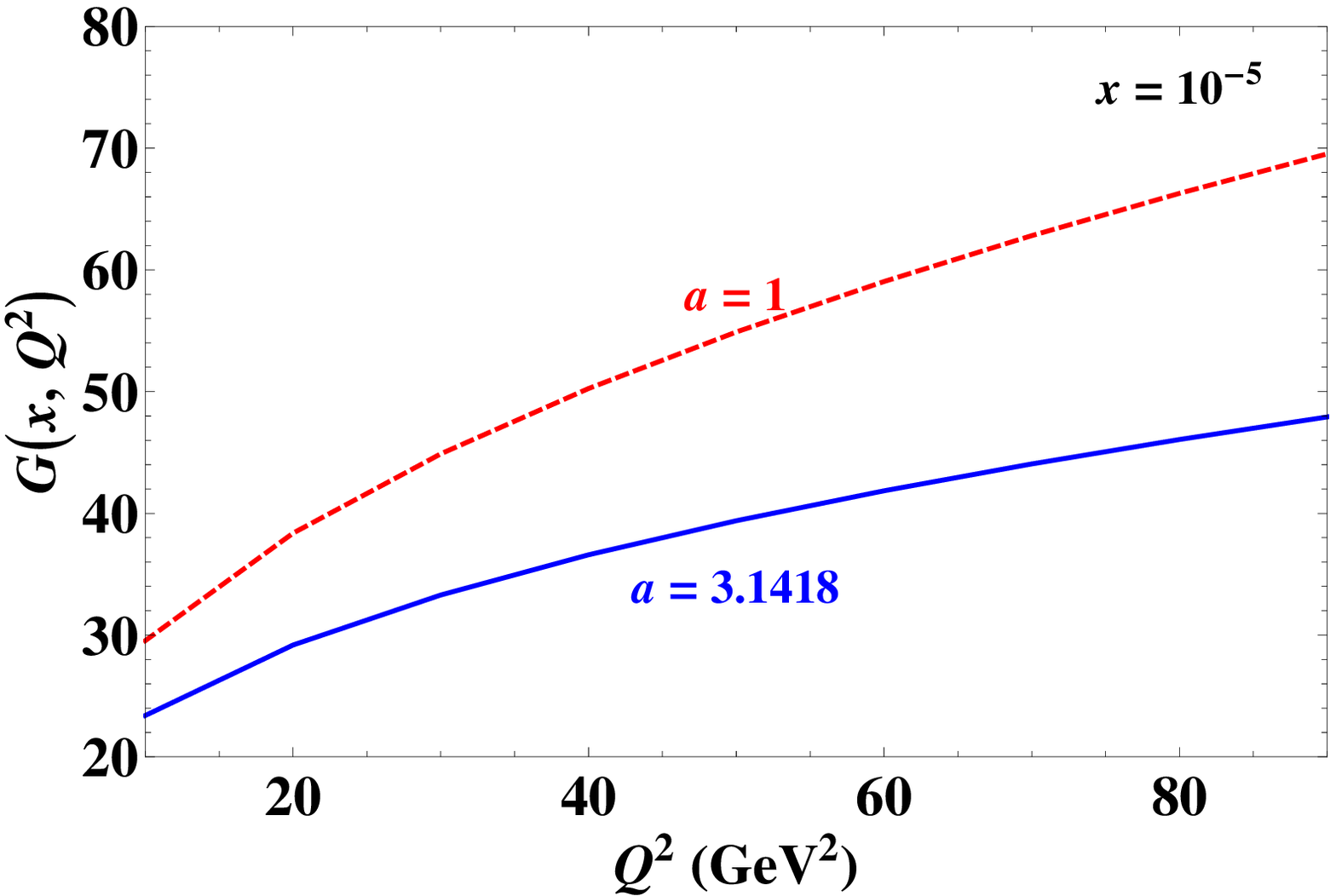}
\caption{The gluon density $G\left(x,Q^{2} \right)$ shown as a function of $ Q^{2} $ at the given values of $x$. The graphs indicate that as \textit{x} decreases, the corresponding gluon density increases rapidly.}
\label{fig:G2}
\end{minipage}
\end{figure}

The above analysis indicates that a generalization of Eq. (\ref{eqn:JKS}) \cite{13} is necessary to be compatible with both momentum analyses of Ref.\cite{8,9} as well as the theoretical expectation of Ref.\cite{14}. Thus, evaluating the value of $F_{2}\left(x,Q^{2} \right)$ (Eq. (\ref{eqn:F2})) and using the corresponding values of $h \left(Q^{2}\right)$ from Table \ref{tab:hQ2}, we obtain the values of $G\left(x,Q^{2} \right)$ for different sets of \textit{x} and $Q^{2}$. Its qualitative feature can be studied from Figure \ref{fig:G1} and Figure \ref{fig:G2}. While in Figure \ref{fig:G1}, we plot $G\left(x,Q^{2} \right)$ vs \textit{x} for a few representative values of $Q^{2}$ = 10, 30 and 60 $\mathrm{GeV^{2}}$, Figure \ref{fig:G2} gives the plot for $G\left(x,Q^{2} \right)$ vs $ Q^{2} $ for \textit{x} = $10^{-3}$, $10^{-4}$ and $10^{-5}$. In both the cases, the dashed curves and the solid curves denote graphs for $ a = 1$ and $ a = 3.1418 $ respectively.\\

\begin{figure}[h]
\mbox{\subfigure{\includegraphics[scale=.5]{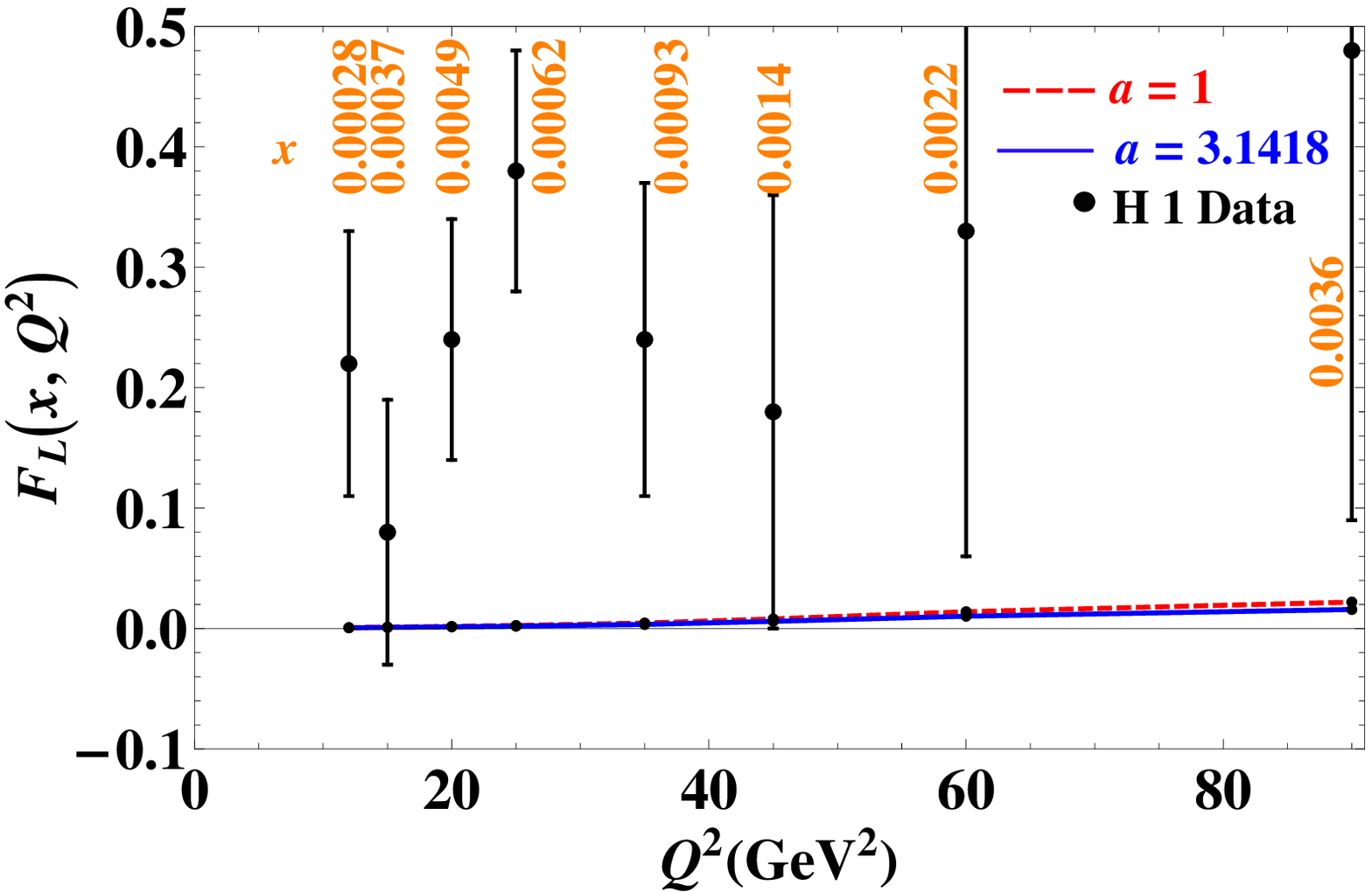}}\,
\subfigure{\includegraphics[scale=.5]{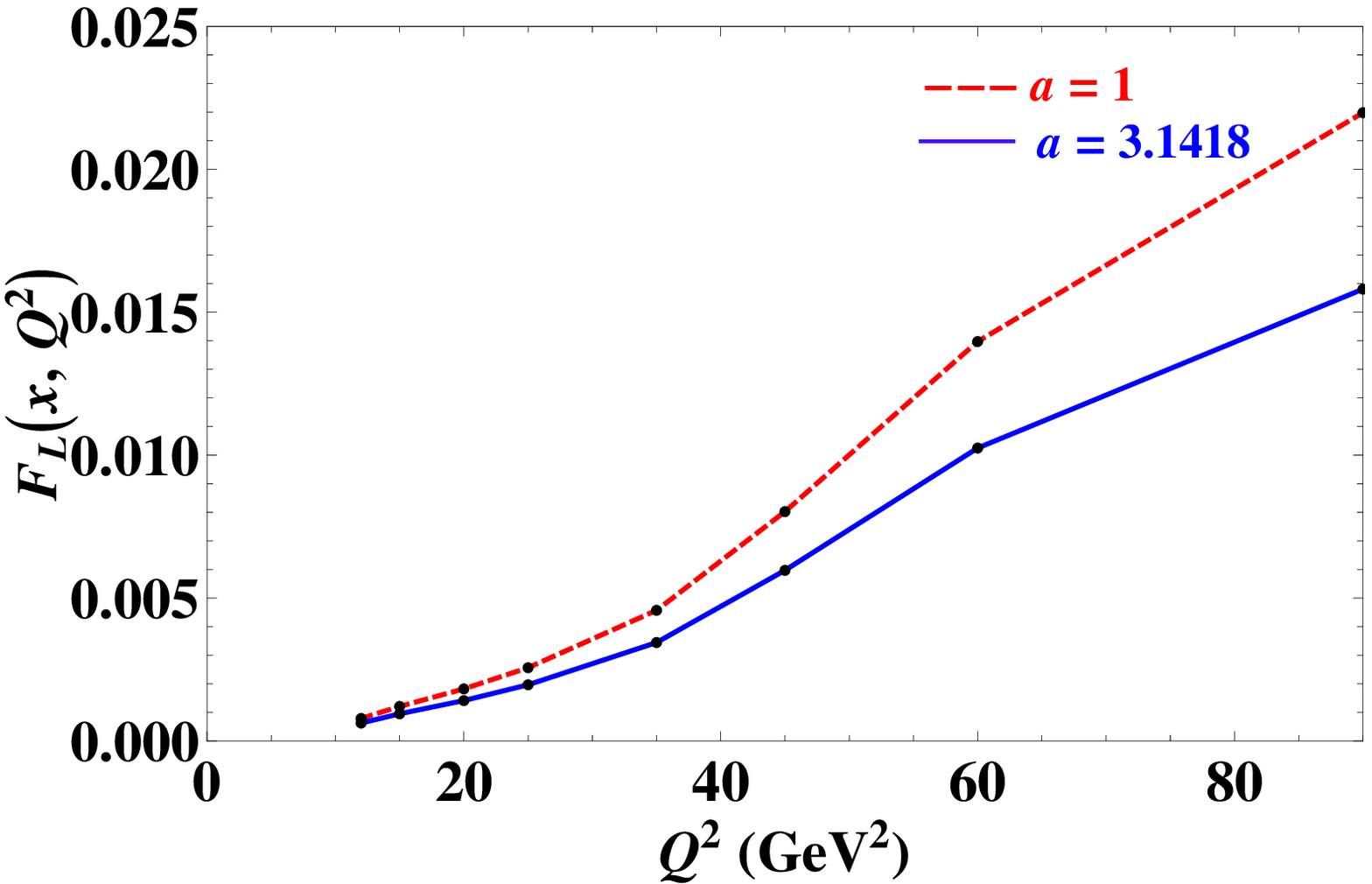}}}
\caption{The longitudinal structure function $F_{L}\left(x,Q^{2} \right)$ shown as a function of $Q^{2}$ at the given values of \textit{x}. The dashed curves describe graph for $a = 1$ and that of solid curves for $a = 3.1418 $. The figure on the right side is the zoomed out image of these two curves (which almost overlap in the left figure).}
\label{fig:FL}
\end{figure}

In Figure \ref{fig:FL}, we plot the predicted lower bound on $F_{L}$ both for $ a = 1 $ (integral charged partons) and $ a = 3.1418$ (fractional charged partons). In the same plot, we also show the available experimental data on $F_{L}$. The data are invariably above the predicted theoretical lower limit as expected.

\section{Conclusions}
In the present work, we have pinned down the gluon distribution function from a self-similarity based model of proton structure function at small \textit{x} using momentum sum rule. We then computed the longitudinal structure function $ \hat{F}_{L}\left(x,Q^{2} \right)$  valid in the same kinematical region as that of the parent model \cite{4}. This results in only partial information of the longitudinal structure function in the form of a lower bound. Compared to our previous analysis \cite{1} on the same, the present one is more reliable since the kinematical region of validity is well-defined. Moreover, we also avoided the problem of singularity of the model at $x \approx 0.019$ which lies outside its phenomenological range of validity. It will be interesting to see the effects of the higher order terms of the expansion of Eq. (\ref{eqn:series}) in the present analysis.\\

The applicability of the method is, however, not limited only to a self-similarity based model. Rather, it can be used in any DGLAP based models as well \cite{18,19} if their range of validity in a specific kinematic range of \textit{x} is well-determined phenomenologically.

\section*{Acknowledgment}
One of the authors (A.J.) thanks the UGC-RFSMS for financial support.

\end{document}